\documentclass[showpacs,preprintnumbers,amsmath,amssymb,prb,twocolumn,floatfix,showkeys]{revtex4}

\usepackage{graphicx}               
\usepackage{psfrag}
\usepackage{dcolumn}                
\usepackage{bm}                     
\usepackage{subfigure}

\newcommand{\ba}{\begin{array}}
\newcommand{\ea}{\end{array}}
\newcommand{\bc}{\begin{center}}
\newcommand{\ec}{\end{center}}

\newcommand{\n}{\Hat{n}}

\begin{document}

\newcommand{\Eq}[1]{{Eq.~(\ref{#1})}}
\newcommand{\EQ}[1]{{Equation~(\ref{#1})}}
\newcommand{\av}[1]{{\left<{#1}\right>}}
\newcommand{\E}{{\textrm{e}}}
\newcommand{\note}[1]{.\newline\marginpar{\LARGE\bf!}{\bf #1}\newline}

\newcommand{\cdag}{c^\dagger}
\newcommand{\cnod}{c^{\phantom{\dagger}}}
\newcommand{\adag}{a^\dagger}
\newcommand{\anod}{a^{\phantom{\dagger}}}
\newcommand{\ctdag}{\tilde c^\dagger}
\newcommand{\ctnod}{\tilde c^{\phantom{\dagger}}}

\title{Ferromagnetic polarons in the one-dimensional ferromagnetic
  Kondo model with quantum mechanical $S=3/2$ core spins}

\author{Danilo R.\ Neuber}
\email{neuber@itp.tu-graz.ac.at}
\affiliation{Institute for Theoretical and Computational Physics,
   Graz University of Technology, Petersgasse 16, A-8010 Graz,
   Austria.}
\author{Maria Daghofer}
\affiliation{Institute for Theoretical and Computational Physics,
   Graz University of Technology, Petersgasse 16, A-8010 Graz,
   Austria.}
\author{Reinhard M.\ Noack}
\affiliation{Fachbereich Physik, Philipps-Universit\"at Marburg,
  D-35032 Marburg, Germany} 
\author{Hans Gerd Evertz}
\affiliation{Institute for Theoretical and Computational Physics,
   Graz University of Technology, Petersgasse 16, A-8010 Graz,
   Austria.}
\author{Wolfgang von der Linden}
\affiliation{Institute for Theoretical and Computational Physics,
   Graz University of Technology, Petersgasse 16, A-8010 Graz,
   Austria.}

\date{October 4, 2005}

\begin{abstract}
  We present an extensive
  numerical study of the ferromagnetic Kondo lattice
  model with quantum mechanical $S=3/2$ core spins. We treat one 
  orbital per site in one dimension using the 
  density matrix renormalization group and include on-site
  Coulomb repulsion between the electrons. 
  We examine parameters relevant
  to manganites, treating the range of low to intermediate doping, 
  $0\alt x<0.5$.
  In particular, we investigate 
  whether quantum fluctuations favor phase separation over the ferromagnetic
  polarons observed in a model with classical core spins. 
  We obtain
  very good agreement of the quantum model with previous results 
  for the classical model, finding separated polarons which are
  repulsive at 
  short distance for finite $t_{2g}$ superexchange $J'$.
  Taking on-site Coulomb repulsion into account, we observe phase
  separation for small but finite superexchange $J'$, while for larger $J'$
  polarons are favored in accordance with simple energy considerations
  previously applied to classical spins.
  We discuss the interpretation of compressibilities and present a 
  phase diagram with respect to doping and the $t_{2g}$
  superexchange parameter $J'$ with and without Coulomb repulsion.
\end{abstract}



\maketitle

\section{Introduction}                                  \label{sec:intro}
The ferromagnetic Kondo lattice model has been widely used as a
minimal model to describe some features of the manganites~\cite{zener51}
La$_{1-x}$Sr$_x$MnO$_3$, 
La$_{1-x}$Sr$_{1+x}$MnO$_4$, and
La$_{2-2x}$Sr$_{1+2x}$Mn$_2$O$_7$. 
The model contains one itinerant $e_g$ orbital and the $t_{2g}$
core spin at each site. 
The $e_g$ electrons are ferromagnetically coupled to the $S=3/2$
core spins generated by the fully occupied $t_{2g}$ orbitals.
The large ferromagnetic Hund's rule coupling
leads to the formation of two bands, the lower and the upper Kondo band,
with the $e_g$ electrons predominantly parallel to the core spins in the lower
band. The core spins strongly influence the mobility of the
$e_g$ electrons via double exchange (DE). At high hole density, this leads
to a ferromagnetic arrangement of the core spins, while
antiferromagnetism is preferred for the completely 
filled lower Kondo band. In the opposite case of an empty Kondo band,
the $t_{2g}$ core spins are antiferromagnetically oriented due to
superexchange.

Since the core spins have a fairly large spin, $S=3/2$, they are
frequently approximated by classical spins, greatly simplifying
calculations. 
Furthermore, the on-site Coulomb repulsion $U$ between the $e_g$
electrons is often neglected because double occupancy is already suppressed
by the Hund's rule coupling and because its
treatment considerably increases the numerical
effort.\cite{KollerPruell2002c}
A review of these semi-classical simulations can be found
in Refs.~\onlinecite{dagotto01:review,furukawa98} and references
therein. Several of these
studies \cite{yunoki98:_phase,dagotto98:_ferrom_kondo_model_mangan}
have found 
that phase separation into regions with either ferromagnetically or
antiferromagnetically  aligned core spins
occurs when the lower Kondo band is nearly empty ($n \gtrsim 0$) or nearly
filled $(n \lesssim 1)$.

In previous work by some of the current authors, also treating the
core spins classically, similar numerical results were 
obtained.\cite{KollerPruell2002c, DaghoferKoller2003}
However, a closer analysis
of the data revealed that the features which had been interpreted by other
authors to indicate phase separation (a discontinuity in the
electron density versus the chemical potential, a pseudogap in the one-particle
density of states) were, in fact, due to small, independent
ferromagnetic polarons.
Likewise, ferromagnetic polarons have been found for the almost empty
lower Kondo band, for localized $S=1/2$ quantum
spins.\cite{Batista_FMPOL98,Batista_FMPOL00} 
For the antiferromagnetic Kondo
model,\cite{Honner_pol} small ferromagnetic droplets were predicted by Kagan et
al.\cite{Kagan_nano}
The question arises whether the correct quantum mechanical treatment of
the $S=3/2$ core spins would favor phase separation instead of
independent ferromagnetic polarons, especially for $T=0$.  
In this paper, we address the influence of quantum spins 
on this issue.

The impact of a quantum mechanical treatment of spins on models for the
Manganites has been addressed in
Refs.~\onlinecite{EdwardsI,Nolting01,Nolting03,Nolting03b}.
In one dimension, quantum mechanical core spins with $S=1/2$ were
employed in a number of studies conducted with the density matrix
renormalization group 
(DMRG).\cite{Batista_FMPOL98,Batista_FMPOL00,Garcia_FMPOL02,Garcia_order00}
Recently, Garcia et al.\cite{garcia_phases04} presented a phase diagram for
$S=1/2$, which, however, was determined for only three values of the density
and did not address the physically interesting region of doping,
$x< 1/3$, treated in this paper.
Quantum mechanical spins with $S=3/2$ have been briefly addressed in an
exploratory study.\cite{dagotto98:_ferrom_kondo_model_mangan} The authors
report phase separation when the lower Kondo band is nearly filled.

In this work, we present extensive calculations for the one-dimensional (1D)
ferromagnetic Kondo lattice model with $S=3/2$ core spins using the DMRG.
We observe that quantum spins yield results in very good agreement with
previous calculations\cite{KollerPruell2002c}  for a model with classical core
spins: for finite $t_{2g}$ superexchange $J'$, polarons are favored over
phase separation. 
In addition, we include an on-site Coulomb repulsion $U$ between the $e_g$
electrons: accordingly, the superexchange parameter $J'$ which favors
antiferromagnetic alignment of the core spins is renormalized and the
effective antiferromagnetic coupling $J_{\mathrm{eff}}$ is weakened. This
leads to an increased polaron size for large $J'$ and phase separation for
small $J'$.

The remainder of this paper is organized as follows.
In Sec.~\ref{sec:model}, we define the model Hamiltonian, for which we
present DMRG results in Sec.~\ref{sec:results}. We discuss the
ground state configurations in Sec.~\ref{sec:ground}. 
In Sec.~\ref{sec:pol_rep} we show that the
polarons are actually repulsive at short distance. We then
discuss the impact of the antiferromagnetic $t_{2g}$ superexchange
$J'$ and of the Hubbard repulsion $U$ and show that quantum
mechanical core spins are very well-approximated by classical spins
(Sec.~\ref{sec:class}). We discuss the transition to the homogeneous
ferromagnetic chain in Sec.~\ref{sec:trans} and present a phase diagram in
Sec.~\ref{sec:phase_diagram}. 
Negative compressibility and the discontinuity in the density vs. the
chemical potential are often taken as a sign for phase separation. In
Sec.~\ref{sec:compr}, we argue that negative compressibility
is an uncertain result when obtained from numerical methods such as the DMRG,
quantum Monte Carlo, or exact diagonalization and show 
that a discontinuity in the density can equally well result from small
independent polarons. 
Finally, Sec.~\ref{sec:conclusion} summarizes and discusses the
results presented in this work. 

\section{Model Hamiltonian and Method}        \label{sec:model}

We study the ferromagnetic Kondo lattice model with localized quantum 
core spins $S=3/2$ and one orbital per site, including a small
Heisenberg-like superexchange between the core spins as well as an on-site
Coulomb repulsion $U$ between the $e_g$ electrons:
\begin{multline}\label{eq_H}
      \hat H = -\sum_{\langle ij\rangle,\sigma}\;
      t\;c^\dagger_{i\sigma}\,c^{\phantom{\dagger}}_{j\sigma}
      + U\sum_{i}\n_{i,\uparrow}\n_{i,\downarrow}\\
      - J_H \sum_{i} \vec{s}_{i} \cdot \vec{S}_i 
      + J'\sum_{\langle ij\rangle} \vec{S}_i \cdot \vec{S}_j\;,
\end{multline}
where $c^{\phantom{\dagger}}_{i\sigma}$ ($c^\dagger_{i\sigma}$) creates
(destroys) an $e_g$ electron with spin $\sigma$ at site $i$,
$\n_{i,\sigma}=c^\dagger_{i\sigma}\,c^{\phantom{\dagger}}_{i\sigma}$
is the corresponding density operator, $\vec{S}_i$ is
the core spin at site $i$ and $\vec{s}_i$ the electron spin. The first term
describes the electron hopping; the hopping integral $t=1$ will be
used in the following as the unit of  
energy. The second term describes the Coulomb repulsion for the $e_g$
electrons; we will treat $U=0$ and $U=10$. 
The third term describes the ferromagnetic Hund's rule coupling
between the $e_g$  electrons and the $t_{2g}$ core spins. 
In this work, we take $J_H=8$, which corresponds to $J_H=6$ in
Refs.~\onlinecite{KollerPruell2002a} and \onlinecite{KollerPruell2002c}, if one
compensates for the normalization of classical core spins to $|\vec{S}|=1$.
The last term describes an additional direct superexchange between the core
spins. For manganites, this effective interaction favors antiferromagnetic
ordering of the core spins, i.e., $J'>0$. We vary $J'$ from
$J'=0$ to $J' = 0.02$. 
Note that $J'=0.01$ corresponds to $J'=\frac{9}{4}\cdot 0.01 \approx 0.02$ in
the units of Refs.~\onlinecite{KollerPruell2002a} and
\onlinecite{KollerPruell2002c}. 

We employ the density matrix renormalization group, 
keeping up to 1000 states at each DMRG iteration and treating chains of
length $L=24$ with open boundary conditions.  The discarded weight
is at most of order $10^{-6}$ for the results presented here. 
Our calculations indicate, however, that 48 states, as used in
Ref.~\onlinecite{dagotto98:_ferrom_kondo_model_mangan}, would lead to
insufficient accuracy for the system sizes treated here.

\section{DMRG Results}                           \label{sec:results}

\subsection{Ground state configurations}\label{sec:ground}

\subsubsection{Polarons are repulsive}\label{sec:pol_rep}

\begin{figure}[hbt]
  \centering
    \includegraphics[width=0.45\textwidth]{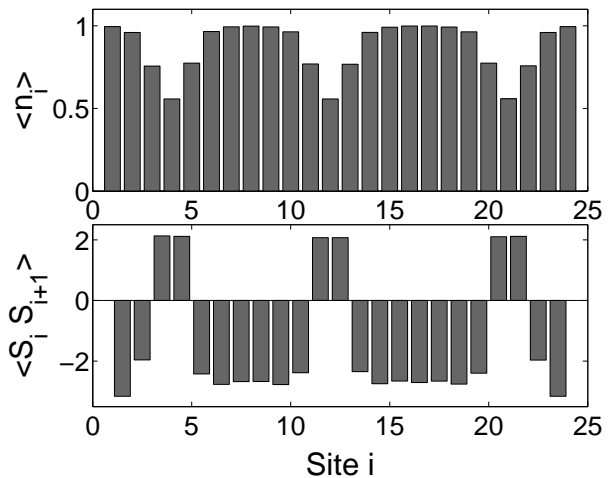} \\
    \caption{Local density $\langle n_i \rangle$ and core spin-core spin
      correlations $\langle S_i S_{i+1}\rangle$ of the DMRG ground state
      for a chain of length $L=24$ with $J'=0.01$ and $U=0$. 
      Three well-separated polarons can be clearly seen.}
    \label{fig:L24N21Jh0.01U0}
\end{figure}

Figure~\ref{fig:L24N21Jh0.01U0} shows the on-site electron density and
the nearest-neighbor core spin correlation
$\langle S_iS_{i+1}\rangle$ for 
a DMRG ground state for a chain of
length $L=24$ for $J'=0.01$ and $U=0$ with three holes. 
The holes clearly form three individual polarons, each
extending over approximately 
three sites with ferromagnetically aligned core spins that are 
embedded in an antiferromagnetic background.

Since the DMRG is a variational method,
convergence to the true ground state is not guaranteed. 
Therefore, we have checked the consistency of the ground state by trying 
different system buildup strategies with respect to particle injection for
many parameter sets. 
Thereby, we have examined whether there is degeneracy of states with
respect to polaron positions and how far the existence of well-separated
polarons and their positions are determined by the details of the DMRG
algorithm.
In addition, one can perform calculations with different $z$-components
$S_z^{\mathrm{tot}}$ of the total spin $S^{\mathrm{tot}}$: increasing
$S_z^{\mathrm{tot}}$ reduces the size of the Hilbert space. This, in general,
allows for more accurate results, while the smaller Hilbert space still
contains the ground state as long as $S_z^\mathrm{tot}\leq S^\mathrm{tot}$.

In an $L=24$ system with two holes ($J'=0.01,\;U=0$), we always find two 
polarons regardless of how the system is built up. 
All energies obtained are degenerate within the estimated numerical
accuracy. 
The polaron position, however, depends on where the particles are injected.
In all cases, we find polarons that are separated by at least a few
lattice sites with 
AFM order. Even if we add both holes simultaneously during the system
buildup, they separate into two polarons. 
Therefore, we conclude that well-separated polarons are effectively
independent, while they seem to repel each other at short distances.

\begin{figure}
  \includegraphics[width=0.4\textwidth]{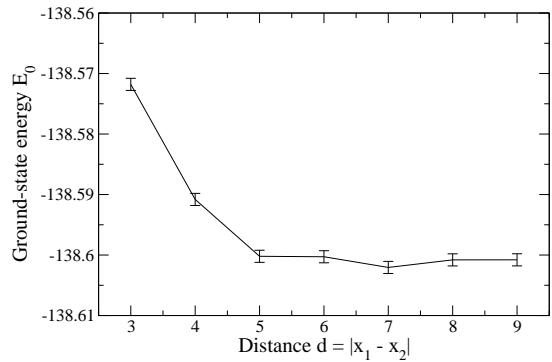}
  \caption{Ground-state energy $E_0$ as a function of the distance $d$ of 
    two polarons on an $L=24$-site chain with $J'=0.01, U=0$. 
    The holes are centered at the position of electrostatic impurity
    potentials $E_{\mathrm{pot}} = -0.1$ 
    and exhibit a symmetric density distribution.
    The energies for $d\geq 5$ are degenerate within the estimated numerical
    accuracy as designated by the error bars.
    \label{fig:Epot}}
\end{figure}

We estimate the energy connected to this repulsion 
by introducing small electrostatic potentials $E_{\mathrm{pot}}=-0.1$ in
order to trap the holes at sites $x_1$ and $x_2$. 
For all distances $d = |x_2 - x_1| \geq 3$, we obtain FM polarons 
covering three lattice
sites with on-site densities symmetric with respect to $x_1$ and $x_2$. 
The ground-state energy as a function of the distance $d=|x_2-x_1|$ is shown
in Fig.~\ref{fig:Epot}, which corroborates the fact that polarons separated
by two or more sites are effectively independent. The configuration with only
one intermediate site is $\Delta E \approx 0.01$ higher in energy, and
in order to obtain a state with adjacent polarons $\Delta E \approx 
0.03$ has to be paid.

We therefore conclude that the holes actually have a \emph{repulsive}
interaction at short range and that separated polarons are energetically
favored  over phase separation into larger FM and AFM regions near the
half-filled Kondo band for $J'=0.01$ and $U=0$. 

The fact that we can obtain degenerate states with a localized polaron at
different positions upon doping with one hole suggests
that polarons are quasi-particles having a large effective mass. 
For physical reasons, we would expect that a single polaron 
delocalizes, forming a heavy Fermi liquid, but it appears that polarons have a
band-width that is too small to be resolved by the DMRG.

\subsubsection{Influence of $J'$ and $U$ and comparison with classical spins}
\label{sec:class}

In Ref.~\onlinecite{KollerPruell2002c},
the energy for phase
separation was compared to that for independent polarons as a function of an
effective superexchange $J_\textrm{eff}$ 
of classical core spins for the almost filled lower Kondo band ($n\lesssim
1$). At small values of
$J_\textrm{eff}$, an increase in the optimal polaron size was
found and bipolarons and phase separation dominated. 
The effective exchange $J_\textnormal{eff}$ is defined in terms of the $t_{2g}$
superexchange $J'$ and the energy $\varepsilon_\textrm{ex}$ for
virtual excitations into the upper Kondo band:
\begin{equation}
  J_{\textnormal{eff}}\approx
  J' + \tfrac{t^2}{\varepsilon_{\textrm{ex}}}\;.
\end{equation}
Without Coulomb repulsion, $\varepsilon_\textrm{ex}$ is given by
the energy for the low-spin state $^3E$.\cite{oles02}
For finite Hubbard $U$, the energy for the virtual excitations is
higher, however, because they lead to doubly occupied sites for $n \approx 1$;
$\varepsilon_\textrm{ex}$ is then given by the energetically higher
$^4E$ and $^4\!A_2$ states.\cite{feiner99} 
Taking into account Coulomb repulsion
should therefore have an effect similar to reducing $J'$.

\begin{figure}
  \includegraphics[width=0.4\textwidth]{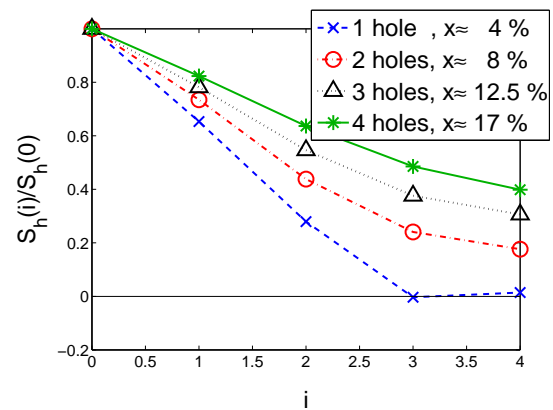}
  \caption{(Color online) Dressed core spin correlation function,
    \Eq{eq:hss}, for classical core spins with effective spinless 
    fermions for $J'=0$, $U=0$, inverse temperature $\beta=100$ and an
    $L=24$-site chain. The ferromagnetic area grows with doping,
    indicating phase separation.\label{fig:hss_classical}} 
\end{figure}
 
\begin{figure}
  \subfigure{\includegraphics[width=0.235\textwidth]{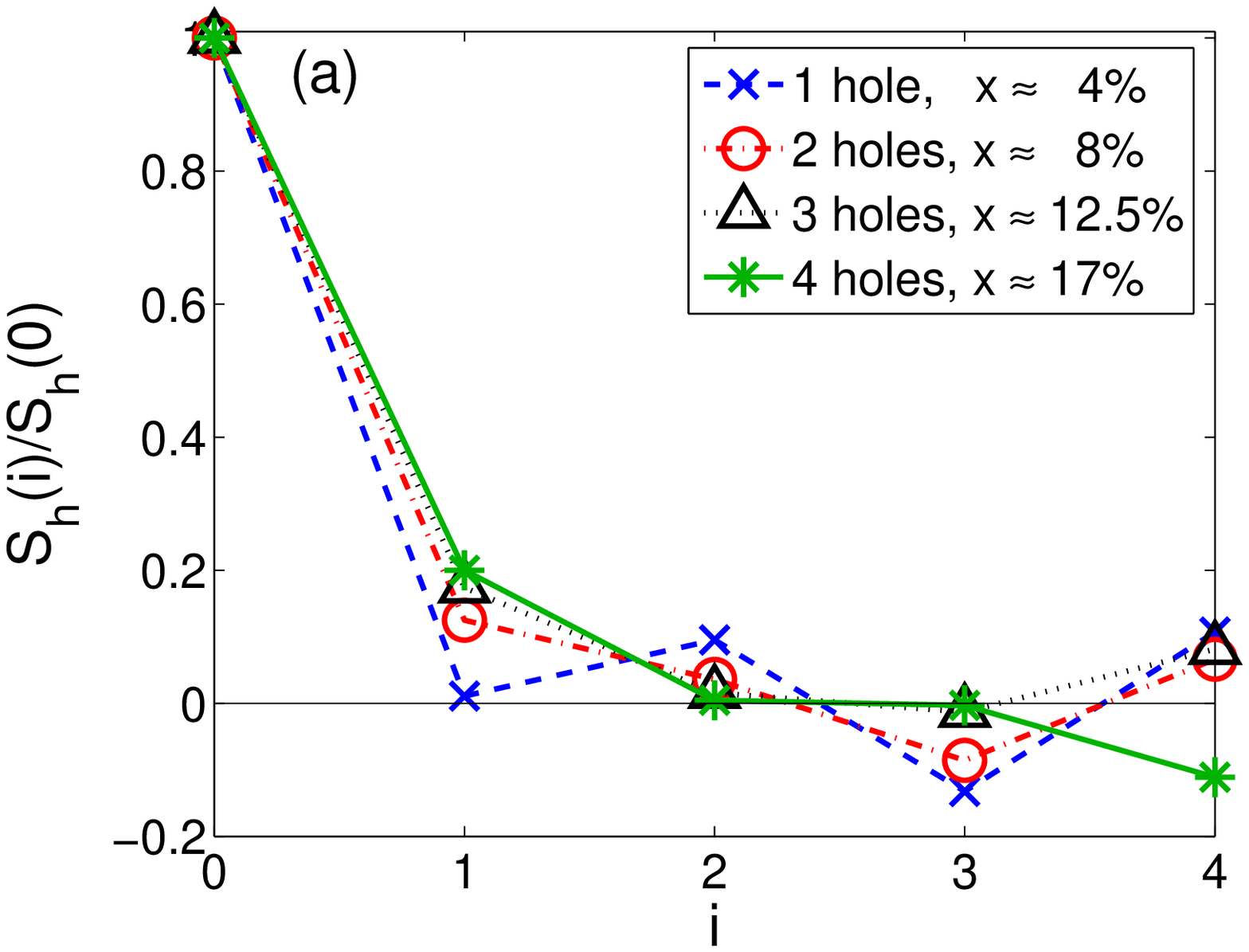}\label{fig:hss_qm_J0.01U0}}
  \subfigure{\includegraphics[width=0.235\textwidth]{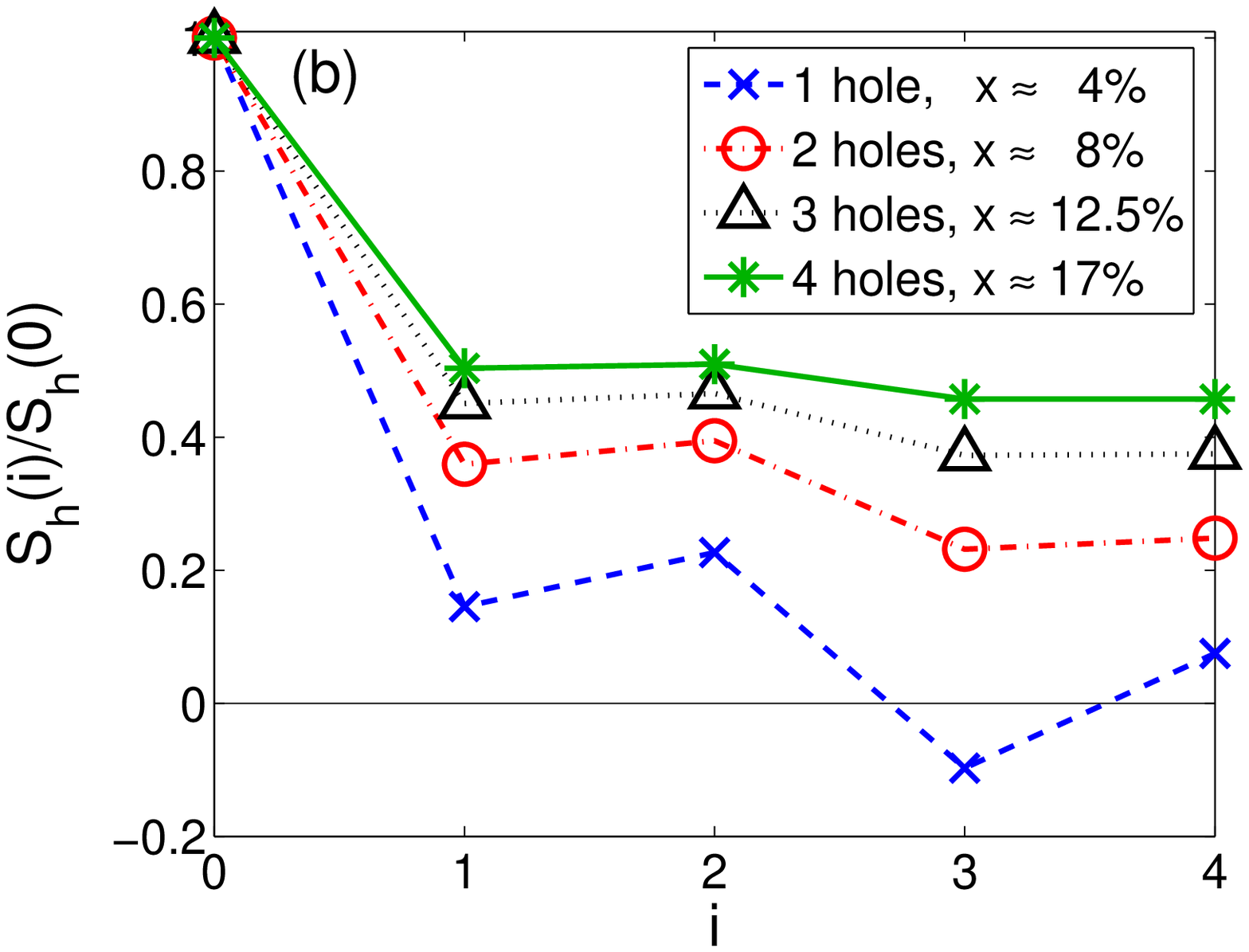}\label{fig:hss_qm_J0.00U0}}\\[-1em]
  \caption{(Color online) Dressed core spin correlation, \Eq{eq:hss},
  for quantum mechanical core spins for $U=0$ and 
  (a) $J'=0.01$ and (b) $J'=0$ on an $L=24$-site chain. 
  In the polaronic regime (a) there is no dependence of the size of the
  ferromagnetic area on doping, in contrast to the phase separated case (b). 
  \label{fig:hss_qm_U0}} 
\end{figure}
 
The spinless fermion model with classical core spins\cite{KollerPruell2002a} 
forms polarons for $J'=0.01$ (corresponding to $J'\approx 0.02$ in the units 
of Ref.~\onlinecite{KollerPruell2002c}), but for $J'=0$ the polarons tend to 
attract each other and phase separate.
This can be seen by examining the dressed core spin correlation
function
\begin{equation}\label{eq:hss}
  S_h(r) = \frac{1}{L-r}\sum_{i=1}^{L-r} n^h_iS_iS_{i+r}\;,
\end{equation}
where $n_i^h=(1-n_{i\uparrow})(1-n_{i\downarrow})$ gives the hole
density relative to the half filled chain, i.e., it is only nonzero
if the site is unoccupied. We have evaluated this observable for
the classical model with $J'=0$. The result, depicted
in Fig.~\ref{fig:hss_classical}, shows that the ferromagnetic regions
around the holes grow with doping, i.e., that the polarons attract each other
and tend to phase separate. 

The dressed core spin correlation function $S_h(r)$ is shown for quantum
mechanical core spins and $U=0$ in Fig.~\ref{fig:hss_qm_U0}.  
For $J'=0.01$ the size of the ferromagnetic regions around the holes
does not grow with doping 
in accordance with our analysis in Sec.~\ref{sec:pol_rep}.
In the case of $J'=0$, we observe phase separation also for the quantum model,
which is reflected in the pronounced increase of FM correlations with doping
in  Fig.~\ref{fig:hss_qm_J0.00U0}. Without an on-site Coulomb repulsion
$U$, we find that the spinless fermion model\cite{KollerPruell2002c} 
agrees qualitatively 
very well with the present DMRG results, confirming that localized $S=3/2$
spins are well-approximated by classical spins. 

If ones takes a closer look at the
numerical values of the dressed core spin correlation function for both
models, one has to keep in mind that states with double occupation have been
projected out in the classical case, while such states are included in the
DMRG calculations. 
In the latter case, there is a finite hole density $\langle
n^h_i\rangle \approx 0.014$ at sites $i$ within the AFM background far
from polarons, and still higher hole densities can be observed 
in the proximity of polarons.  
These contributions of the AFM background to the dressed core spin
correlation function \eqref{eq:hss} partly cancel the FM terms from within
the polarons in the quantum model; this accounts for the almost vanishing
nearest-neighbor correlation $S_h(1)/S_h(0) \approx 0$ for one hole shown in 
Fig.~\ref{fig:hss_qm_J0.01U0}. 

\begin{figure}
  \subfigure{\includegraphics[width=0.235\textwidth]{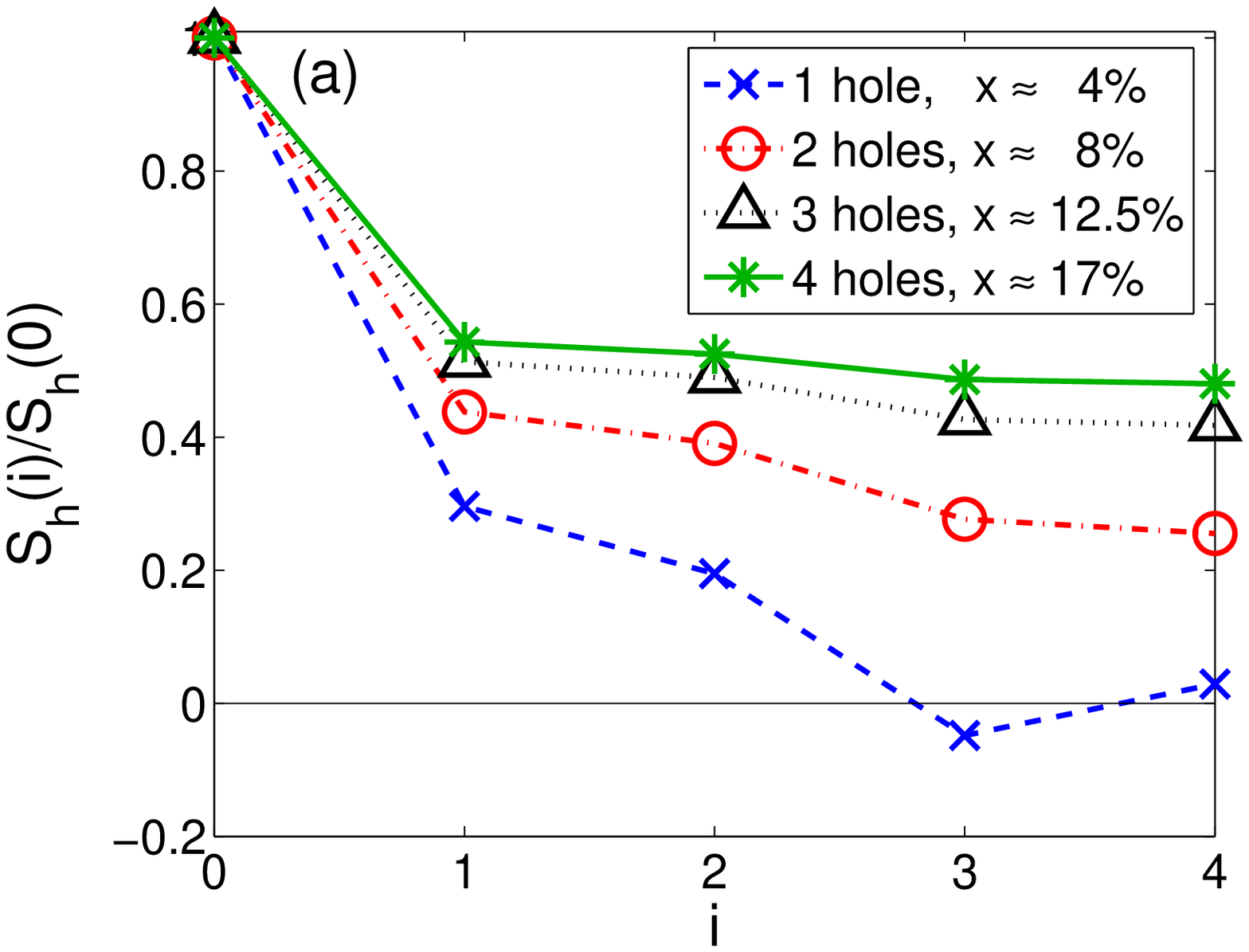}\label{fig:hss_qm_J0.01U10}}
  \subfigure{\includegraphics[width=0.235\textwidth]{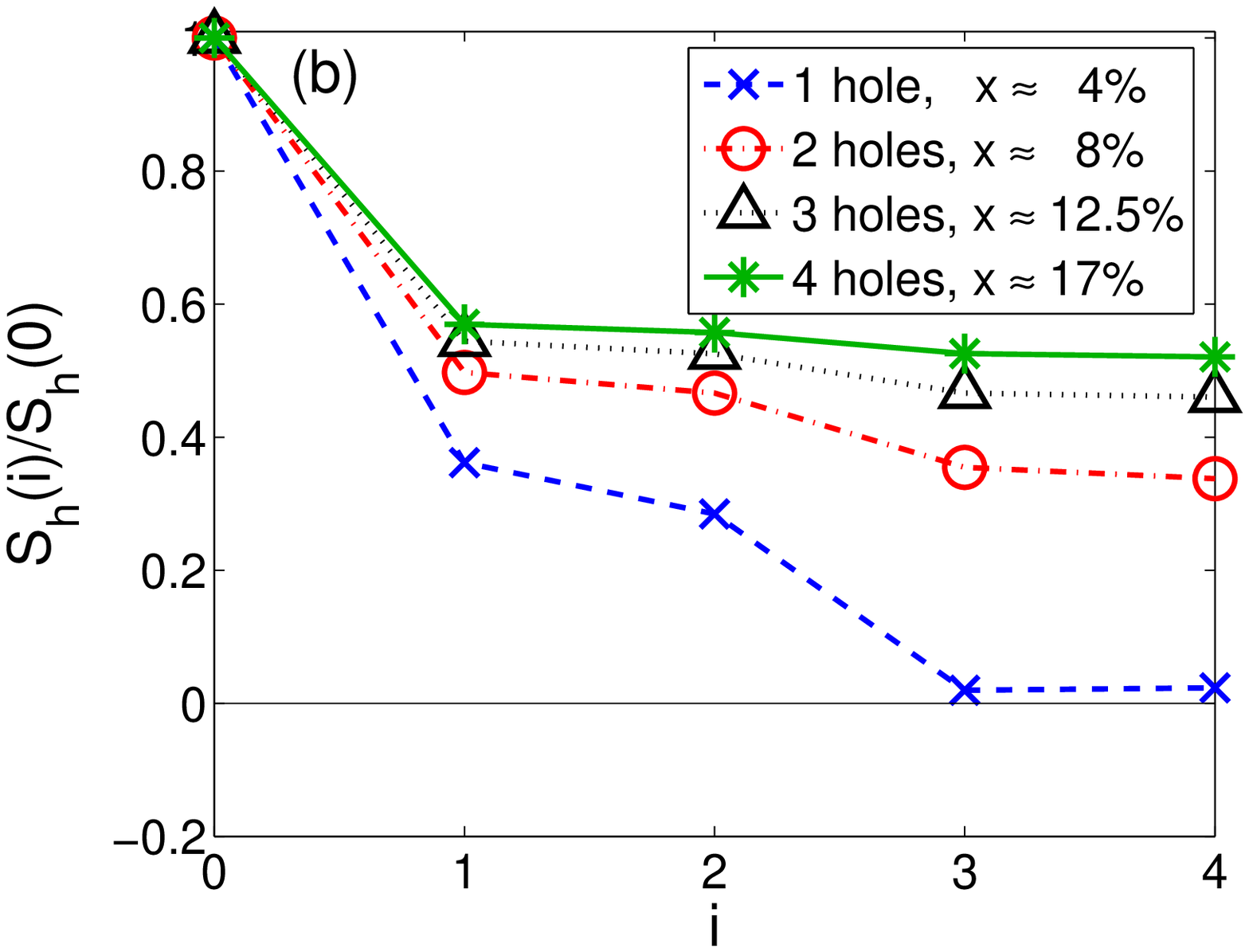}\label{fig:hss_qm_J0.00U10}}\\[-1em]
  \caption{(Color online) Dressed core spin correlation, \Eq{eq:hss}, for
  quantum mechanical core spins for $U=10$ and 
  (a) $J'=0.01$ and (b) $J'=0$ on an $L=24$-site chain. The pronounced
  dependence on doping indicates phase separation.
  \label{fig:hss_qm_U10}} 
\end{figure}

The inclusion of a finite Hubbard $U=10$ decreases the effective exchange
$J_{\mathrm{eff}}$ as discussed above and
therefore increases the tendency to phase separation. 
In fact, phase separation takes place even for $J'=0.01$ 
which is reflected in the dressed core spin 
correlations in Fig.~\ref{fig:hss_qm_U10}.

\subsubsection{Transition to the homogeneous ferromagnetic phase}
\label{sec:trans}

\begin{figure}[htb]
  \centering
  \subfigure[]{
    \includegraphics[width=0.22\textwidth]
    {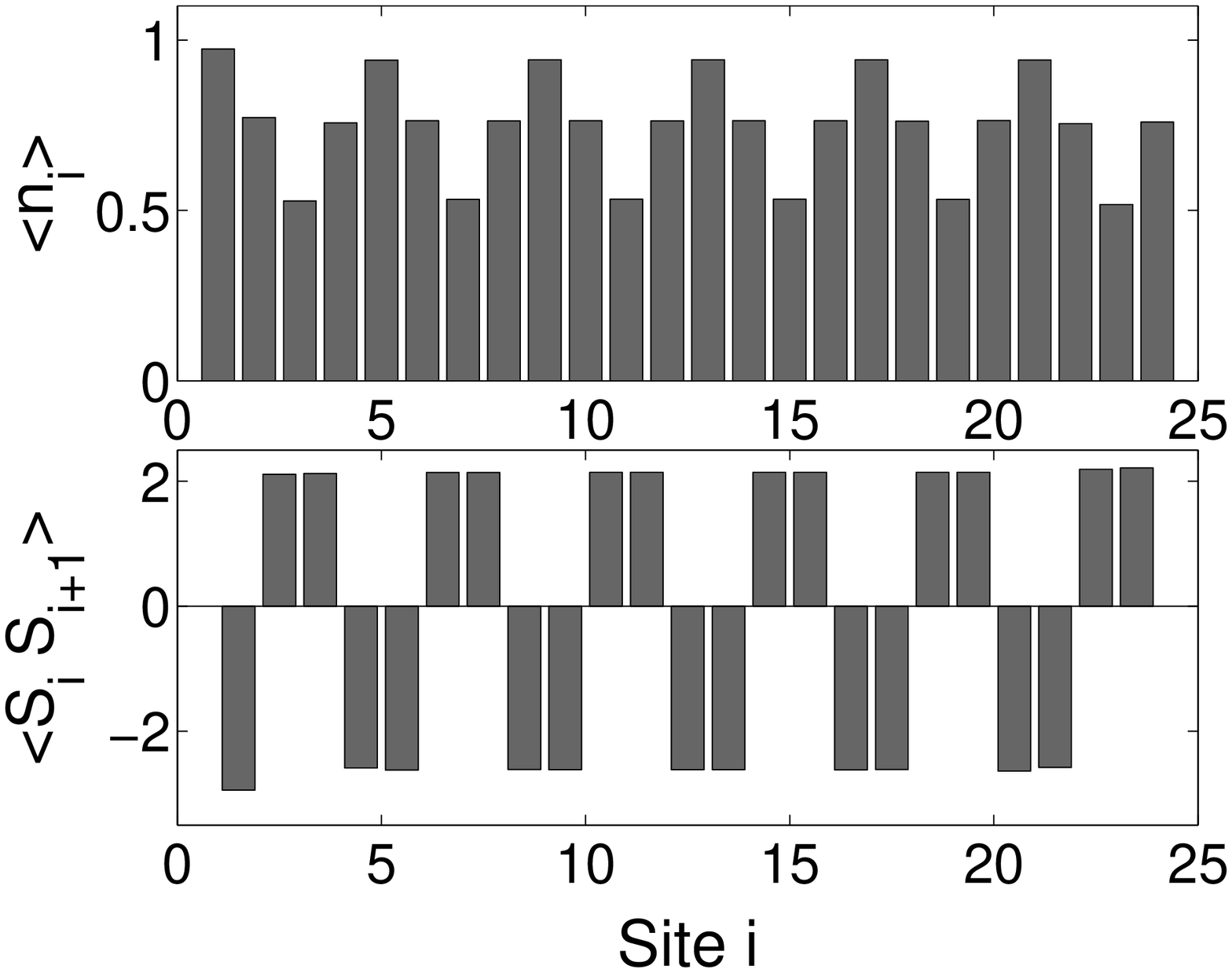} 
    \label{fig:L24N18Jh0.010U0}
  }
  \subfigure[]{
    \includegraphics[width=0.22\textwidth]
    {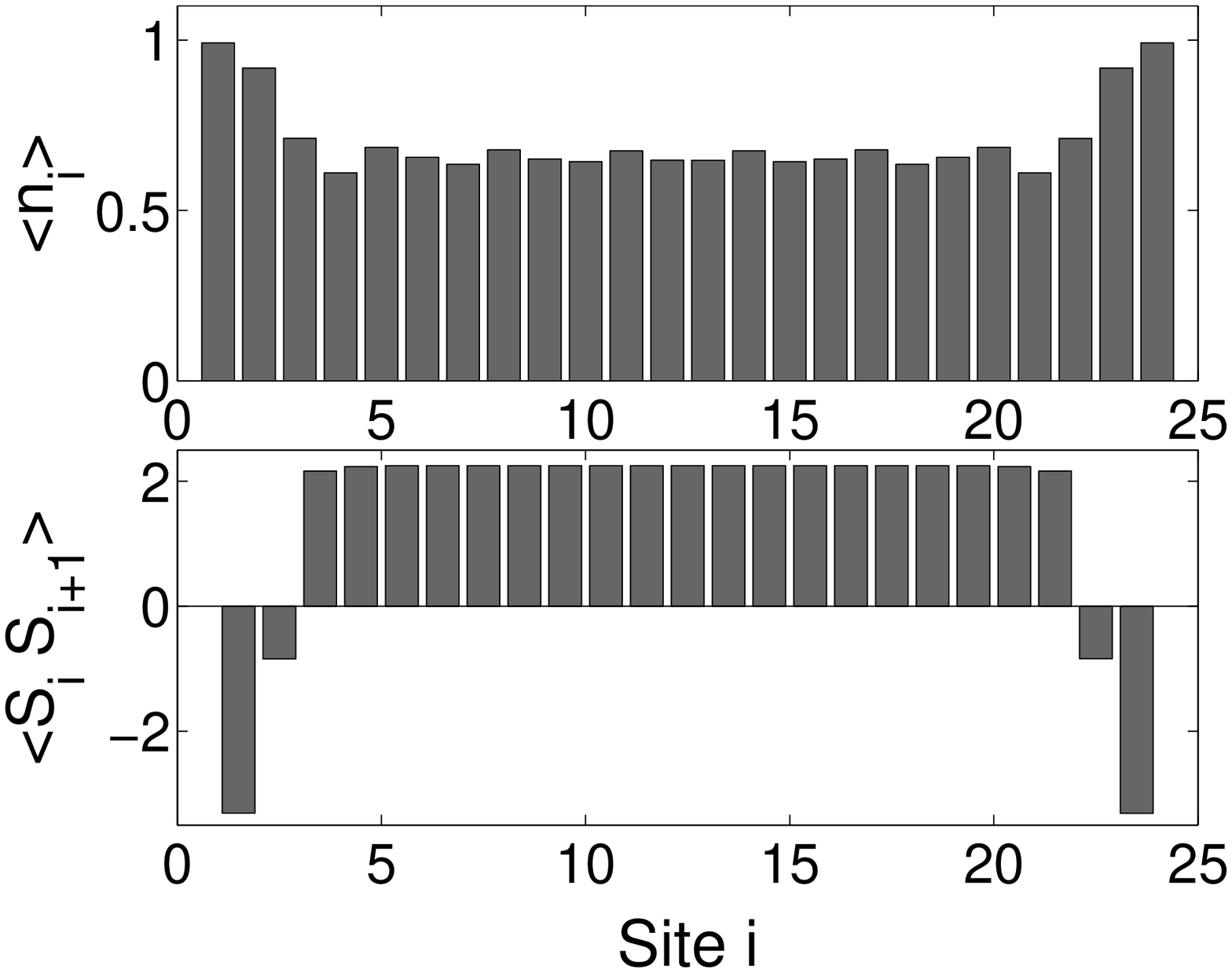} 
    \label{fig:L24N17Jh0.010U0}
  }
  \caption{On-site density $\langle n_i \rangle$ and core spin-core spin
    correlations $\langle S_i S_{i+1}\rangle$ of the DMRG ground
    state for $J'=0.01$, $U=0$ and (a) 6 holes ($x=1/4$) and (b) 7
    holes.
}
\end{figure}

We first examine the polaronic case $J'=0.01$ and $U=0$, where
the polarons extend over approximately 3 sites. 
The ground state of an $L=24$ chain for $x=1/4$ (6 holes) shown in
Fig.~\ref{fig:L24N18Jh0.010U0} corresponds to a periodic arrangement of
polarons, in agreement with the energy considerations of 
Sec.~\ref{sec:pol_rep}.
Similar `island phases' have also been found for $S=1/2$ core spins at
commensurate fillings:\cite{Garcia_FMPOL02} They consist of small
ferromagnetic islands that are aligned antiferromagnetically or have one
anti-aligned spin between them. 
Adding one more hole, we find that the polaronic configuration is
destroyed and a large FM region forms with two AFM arranged spins at each end,
see Fig.~\ref{fig:L24N17Jh0.010U0}. This could be a sign of phase separation,
but much larger chains would have to be treated in order to clarify this issue.
For dopings of $x = 0.375$ (9 holes on an $L=24$ chain)
 we observe complete FM polarization. 

\begin{figure}[htb]
  \centering
  \subfigure[]{
    \includegraphics[width=0.22\textwidth]
    {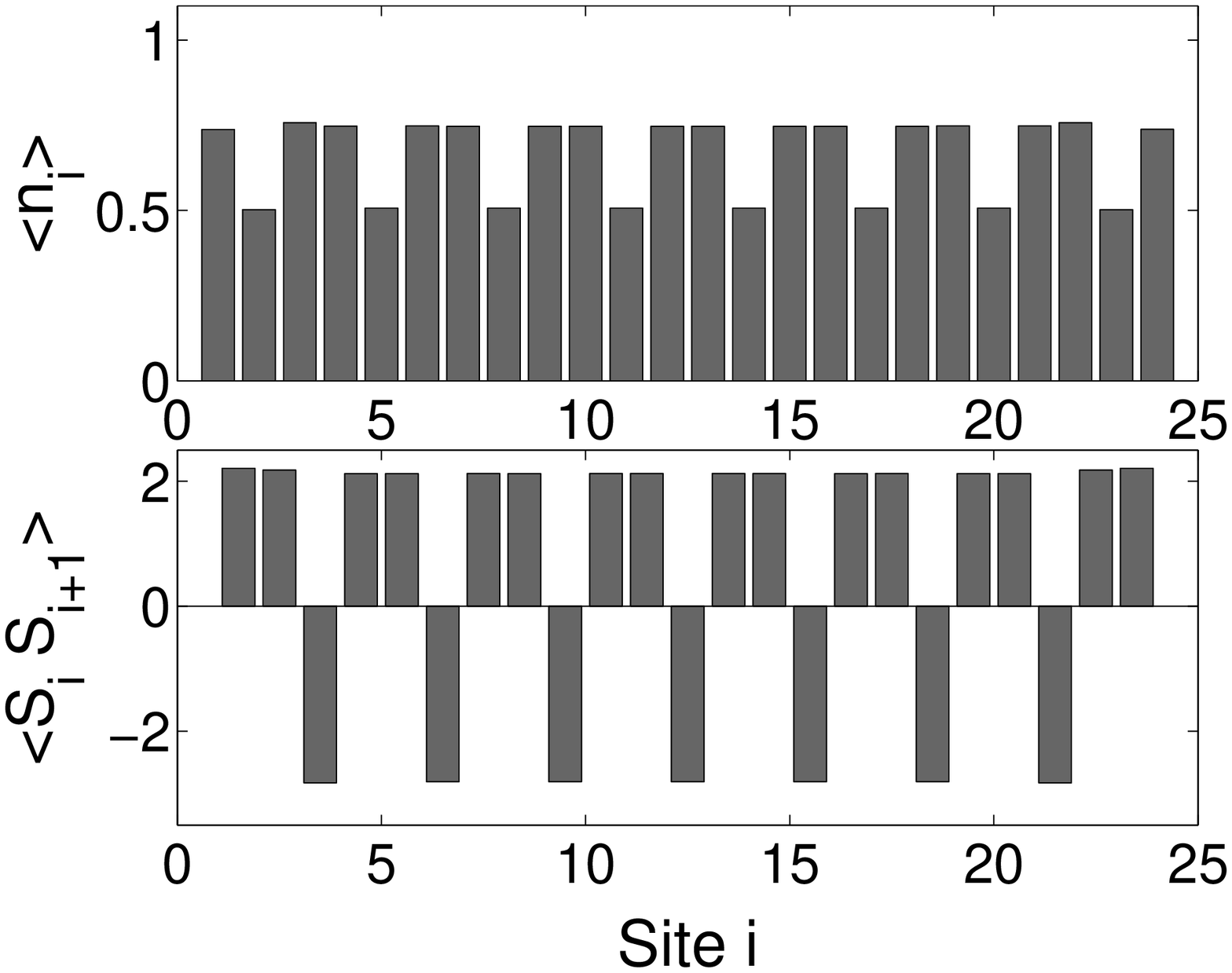} 
    \label{fig:L24N16Jh0.020U0}
  }
  \subfigure[]{
    \includegraphics[width=0.22\textwidth]
    {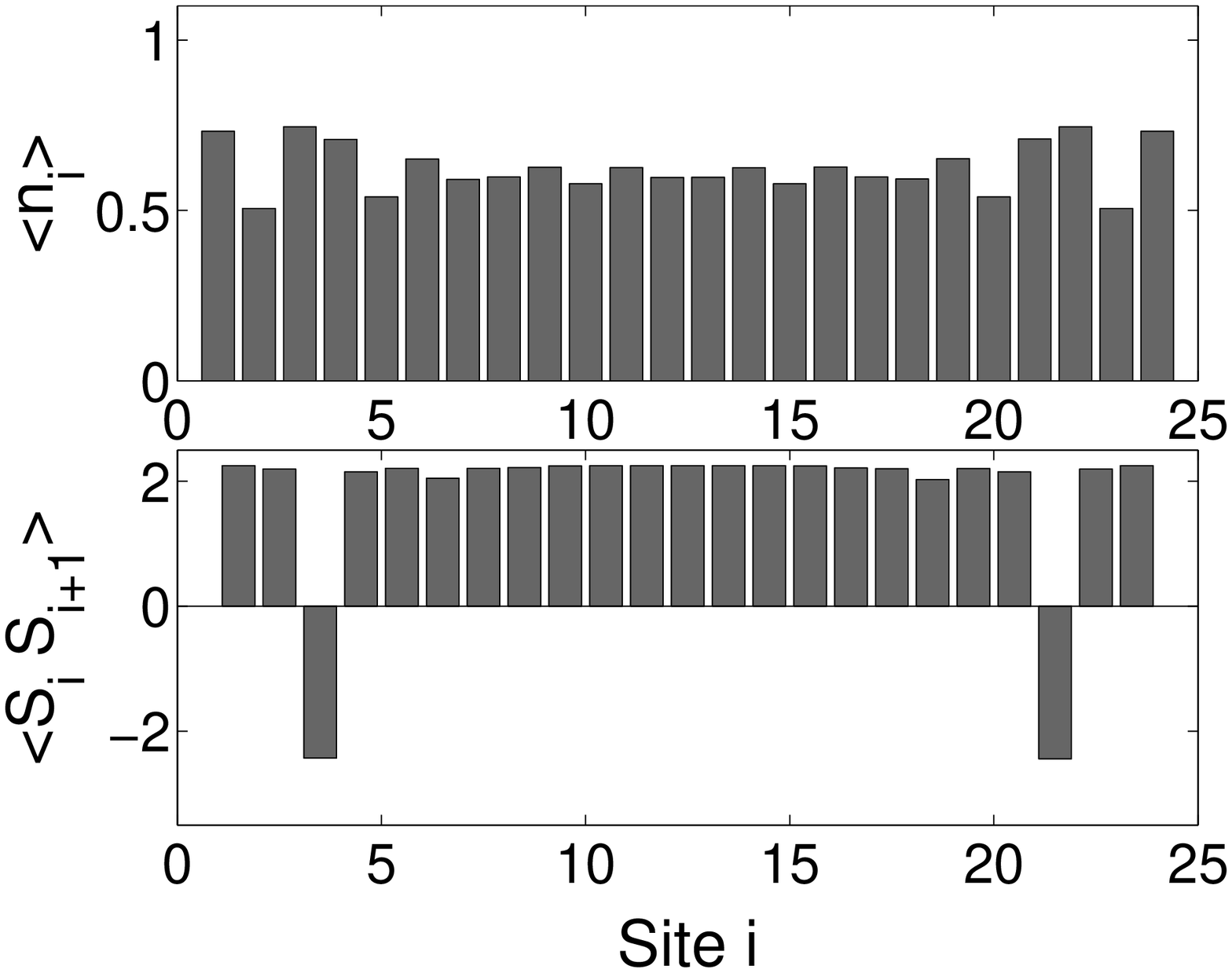} 
    \label{fig:L24N15Jh0.020U0}
  }
    \caption{On-site density $\langle n_i \rangle$ and core spin-core spin
      correlations $\langle S_i S_{i+1}\rangle$ of the DMRG ground
      state for $J'=0.02$, $U=0$ and (a) 8 holes ($x=1/3$) and (b) 9 holes.}
\end{figure}

When the core spin superexchange is increased to $J'=0.02$ (still neglecting
Coulomb repulsion $U=0$), polaronic states become stabilized
up to dopings of $x=1/3$ and we find another island phase of AFM stacked
polarons for 8 holes; see Fig.~\ref{fig:L24N16Jh0.020U0}. 
The result for 9 holes in Fig.~\ref{fig:L24N15Jh0.020U0} suggests phase
separation between an island phase and a FM region; 
however, chains of
length $L=24$ are too small to make definitive statements. As the treatment of
much larger systems is not feasible using currently available computational
resources and the main focus of this paper is the almost-filled 
lower
Kondo band, we did not investigate the nature of the phase transition
further. 
We finally note that upon doping the $L=24$ chain with 10 or more holes, 
complete FM polarization is obtained (not shown).

For parameters $J'=0.02$ and $U=10$, the polaronic regime
extents up to $x=1/4$ (6 holes), so the transition to ferromagnetism is very
similar to the one obtained for $J'=0.01$ and $U=0$. 
We find an
island phase as already shown in Fig.~\ref{fig:L24N18Jh0.010U0} and for 7
holes a configuration similar to the one shown in
Fig.~\ref{fig:L24N17Jh0.010U0}. 
At dopings of 8 holes ($x=1/3$) complete FM polarization is obtained for
$J'=0.02$ and $U=10$.

\subsection{Phase diagrams}
\label{sec:phase_diagram}

\begin{figure}[htb]
  \centering
  \includegraphics[width=0.45\textwidth]{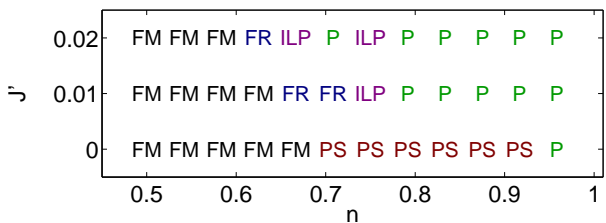}\\
  \caption{(Color online) Phase diagram in the $n$--$J'$ plane for $U=0$. 
    The symbols designate the following characteristics: 
    P: polarons, ILP: island phase (periodic arrangement of polarons), 
    PS: phase separation, FM: ferromagnetic, FR: central ferromagnetic region
    with antiferromagnetically oriented sites at the 
    ends of the chain. The symbols (letters) are determined for an $L=24$-site
    chain. 
  }
  \label{phases_L24U0}
\end{figure}

We summarize the impact of the parameters $J'$ and $U$
in phase diagrams in the $n$--$J'$ plane, where $n=1-x$ 
designates the filling and
$J'$ refers to the $t_{2g}$ superexchange. 
Figure~\ref{phases_L24U0} shows the phase diagram for an $L=24$-site
chain without Coulomb repulsion, $U=0$.
One sees a polaronic region (P) near the filled lower Kondo band, 
including periodic arrangement of polarons (island phase, denoted
as ILP) for commensurate fillings. 
For large $J'>0.01$, this region extends to hole densities
$x\approx1/3$, while phase separation is found for vanishing $J'=0$. 

Some of our results indicate that there might be phase separation for
intermediate fillings between polaronic and FM phases at $J'>0$. 
However, much longer
chains would be needed to clarify this issue; see also our discussion in 
Sec.~\ref{sec:trans}. 
Corresponding states with a large central FM region and either some AFM sites 
or polarons at both ends (cf. Figs.~\ref{fig:L24N17Jh0.010U0} and
\ref{fig:L24N15Jh0.020U0}) have been designated as ``FR'' 
in the phase diagram.
Regions are labeled ``FM'' if \emph{all} nearest-neighbor spin
correlations are positive.  

For $U=0$ and $J'=0$, we find phase separation (PS): 
By varying the system buildup, 
we are able to obtain both polaronic configurations and states 
with a larger FM region embedded in an AFM background;
for these parameters, the phase separated states 
always have a lower energy.

While we expect macroscopic polarization in the thermodynamic limit near the
empty Kondo band, as also observed in 1D 
for the $t-t'-U$ model\cite{daul98,daul00} and a Cu-O chain\cite{guerrero01},
we did not investigate the extent of polarization for finite systems in
detail.  We note that saturation should not be expected for quantum 
spins in the thermodynamic limit according to Ref.~\onlinecite{Nolting03b}.

\begin{figure}[htb]
  \centering
  \includegraphics[width=0.45\textwidth]{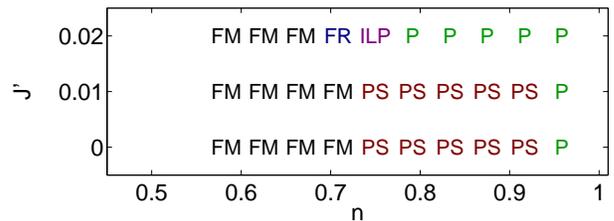}\\
  \caption{(Color online) Phase diagram as in Fig.~\ref{phases_L24U0}, 
    but with $U=10$.\label{phases_L24U10}}
\end{figure}

Fig.~\ref{phases_L24U10} shows the phase diagram for Coulomb repulsion
$U=10$. Comparing with $U=0$, we see that the polaronic phase is suppressed
at lower doping and phase separation takes place even for $J'=0.01$.
This is consistent with the reduction in the effective parameter
$J_\textnormal{eff}$ with increasing $U$.
For parameters $J'=0.01$ and $U=10$, we find that two holes can gain 
$\Delta E \approx 0.008$ forming a bipolaron instead of two separated
polarons.
Compared to the energy $E \approx -137.318$ of the bipolaronic state, 
the energy difference is rather small and this is one reason for the 
sensitivity of the DMRG with respect to system buildup: 
If there are states very close in energy but far apart in phase space,
numerical results will strongly depend on the initial configuration.
We have 
designated the corresponding parts with ``PS'' in
Fig.~\ref{phases_L24U10}. 

\subsection{Compressibility}\label{sec:compr}

The inverse compressibility of a system can be computed approximately 
by numerical differentiation of energies by
\begin{equation}\label{eq:compr}
  \kappa^{-1} = \frac{N_e^2}{L}
  \frac{E(N_e+\Delta,L)+E(N_e-\Delta,L)-2E(N_e,L)}{\Delta^2}\;,
\end{equation}
where $E(N_e,L)$ is the total energy of a chain with $N_e$ electrons on
$L$ sites and $\Delta$ is the difference in particle number.
Negative values are sometimes taken to be an 
indication of phase separation.\cite{dagotto98:_ferrom_kondo_model_mangan}

\begin{figure}[htb]
 \centering
  \subfigure{\includegraphics[width=0.23\textwidth]
    {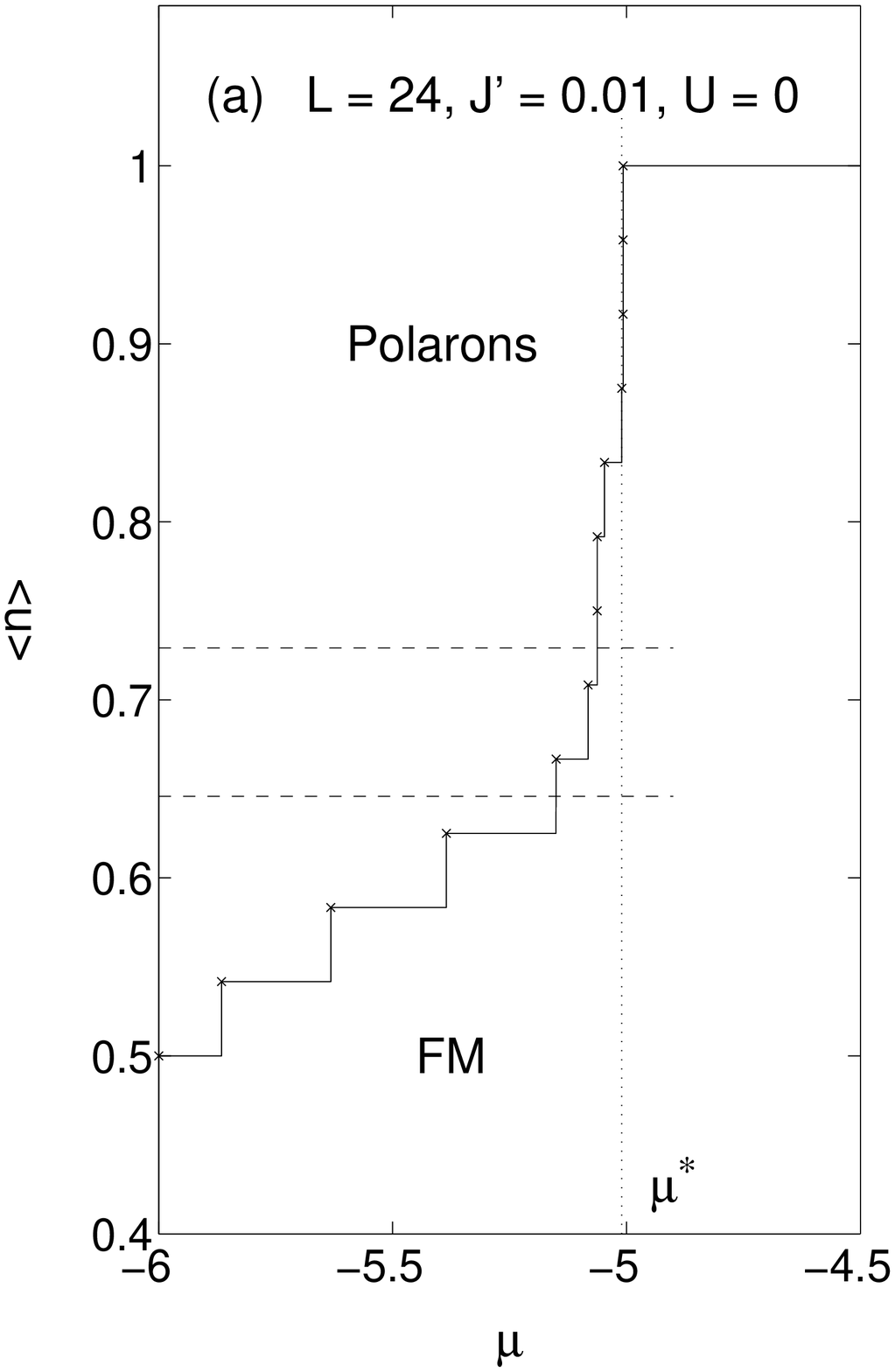}\label{fig:mu_n_U0}}
  \subfigure{\includegraphics[width=0.23\textwidth]
    {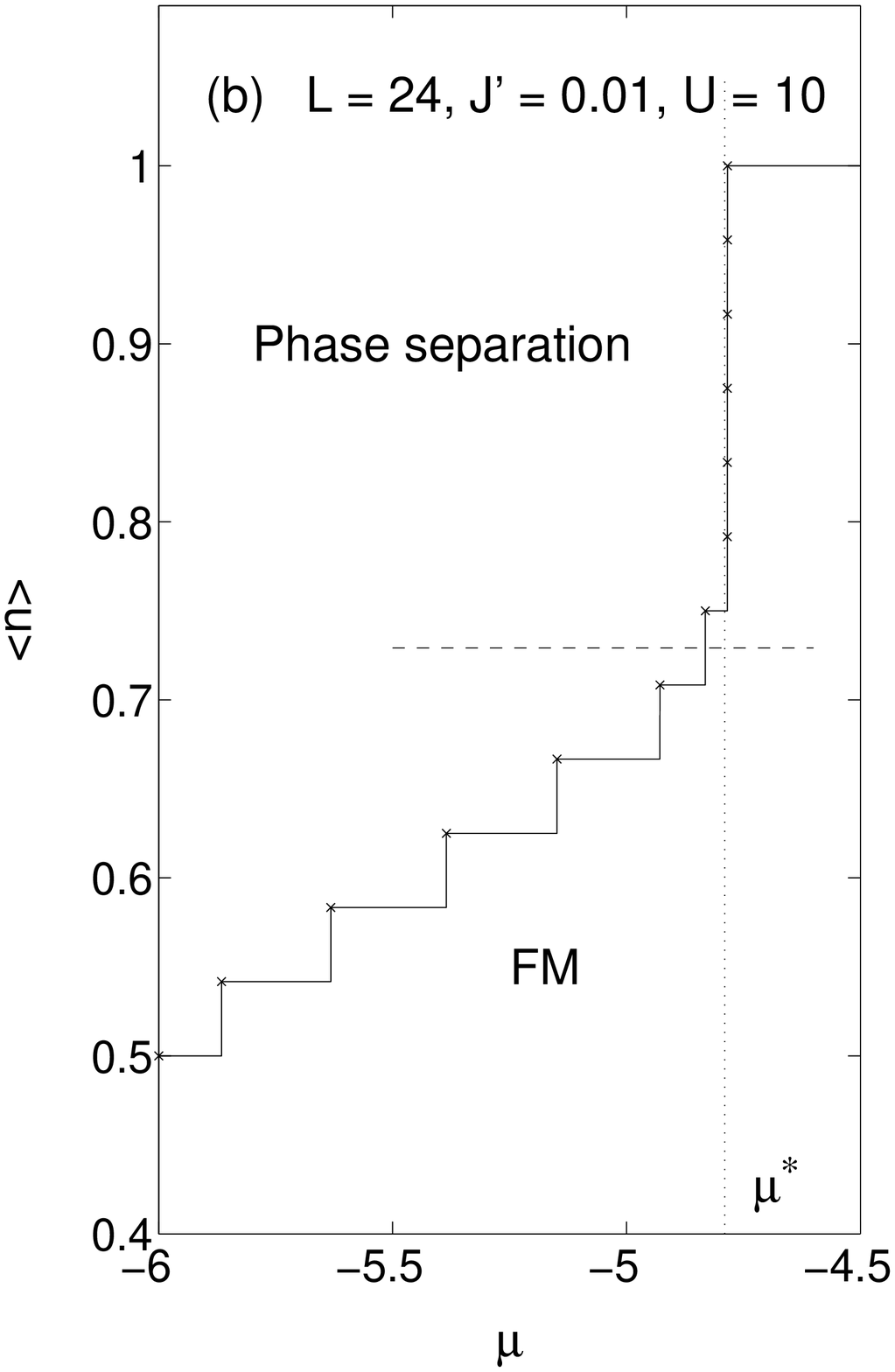}\label{fig:mu_n_U10}}
  \caption[Density vs. $\mu$ for $U=0,\;10$]{Grand canonical
  expectation value  for the electron density 
  $\langle n\rangle$ vs.\ chemical potential $\mu$ for $J'=0.01, L=24$ and
  (a) $U=0$, (b) $U=10$. The dashed lines indicate the limits of the (a)
  polaronic or (b) phase separated and the FM regimes, respectively. 
  \label{fig:mu_n}}
\end{figure}

We want to argue that this condition is neither necessary nor sufficient:
On the one hand, negative values only result from finite-size effects and
should vanish in the thermodynamic limit leading to $\kappa^{-1}\to 0$, as
will be discussed below. 
On the other hand, we will show in Fig.~\ref{fig:mu_n} that polarons can
cause  $\kappa^{-1}\approx 0$ as well. Such
observations therefore have always to be complemented by an investigation of
correlation functions, e.g., the dressed core-spin correlation, 
\Eq{eq:hss}, in order to show the existence of two distinct phases. 

First, we note that numerical methods like the DMRG 
or Monte Carlo calculations 
which do not impose homogeneity, as for instance mean-field theory would,
should yield separated phases in different spatial regions in a proportion
that minimizes the free energy --- in effect, the system performs the Maxwell 
construction by itself and should thus avoid negative compressibilities in
the thermodynamic limit.

For phase separation on finite systems, however, negative compressibilities
$\kappa^{-1}<0$ can indeed arise, because the surface separating the two
phases also contributes to the total energy.  
(From the occurrence of PS, it can be inferred that the phase boundary is not
energetically favorable because the system would otherwise tend to
\emph{maximize} instead of minimizing it and form many small droplets or a
mixed phase.) 
If this boundary has high enough energy and grows with doping, its
contribution to the total energy leads to a negative compressibility. 

In the thermodynamic limit $L\to\infty$, such surface terms become 
negligible compared to the bulk contributions, 
implying $\kappa^{-1} \rightarrow 0$. 
Moreover, the boundary surface does not have to grow with doping at all:
In the present one dimensional case, the phase boundary always consists of
just two bonds connecting the two phases regardless of their size.
For this reason, one could obtain $\kappa^{-1} \approx 0$ here even for
finite systems.  
In addition, one has to keep in mind that numerical differentiation is
notoriously sensitive to even small numerical errors of the ground state
energies entering \Eq{eq:compr}. 

A discontinuity of the density $n$ as function of the chemical potential
$\mu$ is equivalent to the limit of infinite compressibility 
$\kappa^{-1} \rightarrow 0$, and is likewise taken
as an indication for phase separation. We will here show that it can arise
from independent polarons as well.  In order to obtain 
$n(\mu)$ from the DMRG calculations at fixed particle numbers $N_e$, we 
set $n(\mu) = N_e/L$, 
where $N_e$ minimizes the grand canonical expectation
value  $\langle\hat{H}-\mu \hat{N}\rangle = E(N_e,L) - \mu N_e$. 
The results for chains with $L=24,
J'= 0.01, U=0$ and $U=10$, respectively, are shown in
Fig.~\ref{fig:mu_n}. 

We find a jump near $n\approx 1$ both for $U=10$, were PS is observed,
as well as for $U=0$, where we see \emph{polarons}.
In the latter case, it
can be accounted for by the independence of the polarons at low doping,
i.e., adding each polaron costs the same energy, which is 
balanced by the chemical potential $\mu^{\ast} =
\epsilon_{\mathrm{pol}}$.\cite{KollerPruell2002c,DaghoferKoller2003} 
This also corresponds to our observation that the
ground state energy per site as a function of filling (not shown) lies 
practically on an straight line near $n=1$ with $dE/dn =
\epsilon_{\mathrm{pol}}$. 

In conclusion, we have argued that 
a negative compressibility for a finite system is neither a necessary
nor a sufficient condition for phase separation.
It is also important to note that a discontinuity of the
density $n$ as function of the chemical potential $\mu$ is a necessary
condition for phase separation, but is not a sufficient condition
because other mechanisms such
as independent polarons can induce a `jump' of $n(\mu)$ as well.

\section{Conclusions}                           \label{sec:conclusion}
We have presented an extensive 
numerical study of the one-dimensional ferromagnetic Kondo lattice 
model with quantum mechanical $S=3/2$ core spins using the
density matrix renormalization group method,
treating  a non-degenerate conduction band with and without on-site Coulomb
repulsion at low to intermediate doping. In particular, we have explored
the similarities with the analogous model with classical
core spins, where ferromagnetic polarons have 
been found to dominate over phase separation\cite{KollerPruell2002c} for
parameters relevant to manganites.
We have investigated whether the inclusion of
quantum fluctuations leads to an attractive interaction between the polarons
and thus to phase separation. We find 
that this is not the case: the polarons are in fact 
\emph{repulsive} at short distances. 
In general, the results of the quantum model agree very well with the
classical model and show that classical spins are indeed a very good
approximation for manganites.

Furthermore, we have investigated the influence of a relatively large
local Coulomb repulsion $U=10$, which reduces the effective AFM coupling
$J_\textnormal{eff}$ relevant when $n\lesssim1$. For small $t_{2g}$
superexchange $J' \leq 0.01$, 
phase separation dominates over separated polarons
if on-site Coulomb repulsion is taken into account, while polarons are
favored upon increasing the superexchange parameter to $J'=0.02$. 
We have summarized these findings in phase diagrams in the plane of doping 
and $t_{2g}$ superexchange $J'$ for $U=0$ and $U=10$.

We have also argued that the observation of a (small) 
negative compressibility
and of a jump in the density $n$ as a function of the chemical potential
$\mu$ are  not compelling indicators of phase separation. 
While a discontinuity of $n(\mu)$ is a
necessary condition for phase separation, it is not a sufficient one:
different mechanisms such as the formation of independent polarons can leave
similar traces.

\begin{acknowledgments}
This work has been supported by the Austrian Science Fund (FWF), project
no.\ P15834-PHY. 
\end{acknowledgments}

\end{document}